\documentclass[useAMS,usenatbib]{mnras}

\usepackage{graphicx}
\usepackage{color}
\usepackage{amsmath}
\usepackage{amssymb}

\title[HELP: Star formation as a function of galaxy environment with Herschel]{HELP\thanks{\url{herschel.sussex.ac.uk}}: star formation as function of galaxy environment with \textit{Herschel}\thanks{ \url{sci.esa.int/herschel/}}}

\author[S. Duivenvoorden et al.]{S. Duivenvoorden$^{1}$\thanks{E-mail: \href{mailto:S.Duivenvoorden@Sussex.ac.uk}{S.Duivenvoorden@Sussex.ac.uk}},  S. Oliver$^{1}$, V. Buat$^{2}$, B. Darvish$^{3}$, A. Efstathiou$^4$, D. Farrah$^5$, \newauthor M. Griffin$^{6}$, P. D. Hurley$^1$, E. Ibar$^7$, M. Jarvis$^{8,9}$, A. Papadopoulos$^4$, M. T. Sargent$^1$,\newauthor D. Scott$^{10}$, J. M. Scudder$^1$, M. Symeonidis$^{11}$, M. Vaccari$^{9,12}$, M. P. Viero$^{3,13}$, L. Wang$^{14}$\\
$^{1}$Astronomy Centre, Department of Physcis and Astronomy, University of Sussex, Brighton BN1 9QH\\
$^{2}$Aix-Marseille Universite, CNRS, LAM (Laboratoire d' Astrophysique de Marseille) UMR 7326, 13388 Marseille, France\\
$^3$Cahill Center for Astrophysics, California Institute of Technology, 1216 East California Boulevard, MC 278-17, Pasadena, CA 91125, USA\\
$^4$School of Sciences, European University Cyprus, Diogenes Street, Engomi, 1516, Nicosia, Cyprus \\
$^5$Department of Physics, Virginia Tech, VA 24061, USA\\
$^{6}$School of Physics and Astronomy, Cardiff University, Queens Buildings, The Parade, Cardiff CF24 3AA\\
$^{7}$Instituto de F\'isica y Astronom\'ia, Universidad de Valpara\'iso, Avda. Gran Breta\~na 1111, Valpara\'iso, Chile\\
$^8$Astrophysics, University of Oxford, Keble Road, Oxford, OX1 3RH, UK\\
$^{9}$Department of Physics and Astronomy, University of the Western Cape, Robert Sobukwe Road, 7535 Bellville, Cape Town, South Africa\\
$^{10}$Department of Physics \& Astronomy, University of British Columbia, 6224 Agricultural Road, Vancouver, BC V6T 1Z1, Canada\\
$^{11}$Mullard Space Science Laboratory, University College London, Holmbury St. Mary, Dorking, Surrey RH5 6NT\\
$^{12}$INAF - Istituto di Radioastronomia, via Gobetti 101, 40129 Bologna, Italy \\
$^{13}$Kavli Institute for Particle Astrophysics and Cosmology, Stanford University, 382 Via Pueblo Mall, Stanford, CA 94305\\
$^{14}$SRON Netherlands Institute for Space Research, Landleven 12, 9747 AD, Groningen, The Netherlands}
\begin{document}
\date{Accepted 2016 June 16. Received 2016 June 16; in original form 2016 January 21}

\pagerange{\pageref{firstpage}--\pageref{lastpage}} \pubyear{2016}

\maketitle

\label{firstpage}

\begin{abstract}
The Herschel Extragalactic Legacy Project (HELP) brings together a vast range of data from many astronomical observatories. Its main focus is on the \textit{Herschel} data, which maps dust obscured star formation over 1300 deg$^2$. With this unprecedented combination of data sets, it is possible to investigate how the star formation vs stellar mass relation (main-sequence) of star-forming galaxies depends on environment. In this pilot study we explore this question between $0.1 < z < 3.2$ using data in the COSMOS field. We estimate the local environment from a smoothed galaxy density field using the full photometric redshift probability distribution.
We estimate star formation rates by stacking the SPIRE data from the \textit{Herschel} Multi-tiered Extragalactic Survey (HerMES). Our analysis rules out the hypothesis that the main-sequence for star-forming systems is independent of environment at $1.5 < z < 2$, while a  simple model in which the mean specific star formation rate declines with increasing environmental density gives a better description. However, we cannot exclude a simple hypothesis in which the main-sequence for  star-forming systems is independent of environment at $z  < 1.5$ and $z >2$.  We also estimate the evolution of the star formation rate density in the COSMOS field and our results are consistent with previous measurements at $z < 1.5$ and $z > 2$ but we find a $1.4^{+0.3}_{-0.2}$ times higher peak value of the star formation rate density at $z \sim 1.9$. 

\end{abstract}


\section{Introduction}

Galaxies are found in different environments, from rich clusters, to small groups, to isolated galaxies residing in cosmic voids. For several decades there has been substantial evidence that the environment influences galaxy properties, such as star formation rate (SFR), morphology and colour \citep[e.g.][]{1980ApJ...236..351D}.  The environment has been found to influence the timing of the quenching of star formation \citep[e.g.][]{2013MNRAS.432..336W}, though the onset of star formation is not necessarily caused by environment \citep[e.g.][]{2011ApJ...732...94R}.This quenching leads to a decline in the fraction of active star forming galaxies in clusters from $\sim$20 per cent at z $\sim$ 0.4 to almost zero in the local universe \cite[e.g.][]{1984ApJ...285..426B,2013ApJ...775..126H}. 

There are several ways to probe the SFR of a galaxy. For a young stellar population, the radiation is dominated by the UV light emitted from massive stars. Due to the short lifetime of these stars, this emitted power is a good indicator for star formation. However, not all emitted UV light escapes the galaxy. In the presence of dust, a significant percentage of this light is absorbed and reradiated at mid- and far-infrared (FIR) wavelengths. The total FIR luminosity is therefore a function of the amount of obscured UV light, and provides another method to measure the SFR \citep[e.g.][]{1998ApJ...498..541K,2014ARA&A..52..415M}.

Low redshift galaxies in dense environments are redder on average than field galaxies \citep[e.g.][]{2004MNRAS.353..713K,2013MNRAS.434..423K,2013ApJS..206....3S}. This can be partly explained by a higher dust content of galaxies or a lower SFR of individual systems. Although these relations are well established at low redshift, it is still open to debate whether the trends stop, or may even be reversed at higher redshift \citep[e.g.][]{2007A&A...468...33E,2014ApJ...788..133T}.

The environment is not the only parameter that influences the evolution of galaxies. Both internal and external effects determine how much gas is available for forming stars, as the internal gas supply gets replenished by accretion of cold gas from the environment \citep[e.g.][]{2009ApJ...703..785D}. The availability of this cold gas has a significant influence on the SFR of galaxies \citep[e.g.][]{2014A&A...562A..30S}.

To distinguish between environment and internal influences, we need to characterize them simultaneously. Recent studies indicate that quenching is driven more by the internal properties of a galaxy than by the environment \citep[e.g.][]{2015ApJ...806..162H}. However, the fractional role of the internal and external processes in galaxy quenching may depend on, e.g., redshift and stellar mass of galaxies \citep[e.g.][]{2010ApJ...721..193P,2011MNRAS.411..675S,2016arXiv160503182D}. We note that separating the internal and external effects can be difficult because they seem to be strongly connected with each other \citep{2012MNRAS.423.1277D}. 

Another example of the importance of internal galaxy properties on galaxy evolution is the strong relation, between SFR and stellar mass for star-forming galaxies: termed the main-sequence \citep[MS,][]{2004MNRAS.351.1151B}. The MS has been found in both the local Universe and at higher redshifts \citep[e.g.][]{2007ApJ...660L..43N,2014MNRAS.437.3516S,2015MNRAS.449..373D,2015A&A...577A.112L}. A trend is seen for galaxies on the MS, where the more massive a star-forming galaxy is, the higher the star formation rate becomes. In relative terms, the specific star formation rate (sSFR, the star formation rate per unit stellar mass) appears to drop for higher mass galaxies: galaxies with higher total stellar mass are redder and have relatively less star formation per unit mass. The MS seems to be in place out to z $>$ 2.5 \citep{2012ApJ...754L..29W}. However, the specific details of the MS, such as slope and dispersion, vary between different studies \citep[e.g.][]{2014ApJS..214...15S}. Moreover, the sSFR of galaxies on the MS evolves with redshift as roughly $(1+z)^{3}$ out to redshift of 2-3 \citep{2010MNRAS.405.2279O,2015A&A...577A.112L,2015MNRAS.453.2540J,2015ApJ...807..141P}.  

Whether or not the normalisation of the MS depends on environment is still under discussion. \cite{2014MNRAS.445.4045R} found that the MS is constant for void, void-shell, and reference galaxies at $z<0.12$.  \cite{2013ApJ...773...86T,2014ApJ...794...31T} found no difference between the MS in clusters and in the field. \cite{2016ApJ...816L..25P}  found a similar result, but also found a population of galaxies with reduced SFR (departing from the MS) within the virial radius of the cluster at $z \sim 0.1$.

It has been argued that higher density environments only reduce the fraction of galaxies that are star-forming and do not seem to have a major effect on the average SFR of star-forming galaxies \citep[e.g.][]{2010ApJ...721..193P,2012MNRAS.423.3679W,2014ApJ...796...51D,2016arXiv160503182D}. However, at low redshift, \citet[][$z < 0.1$]{2010MNRAS.404.1231V} and  \citet[][$0.15 < z < 0.3$]{2013ApJ...775..126H} found a reduction of the SFR of star-forming cluster galaxies, compared to their counterparts in the field. Furthermore \cite{2012MNRAS.423.2690S} found an enhanced MS in isolated compact groups, but compact groups embedded in larger systems do not have this enhanced SFR.

At intermediate redshift $(0.4 \le z \le 0.8)$ \citet{2010ApJ...710L...1V} found that the SFR of cluster star-forming galaxies was a  factor of 1.5 lower than in the field. This result is in agreement with \cite{2011ApJ...735...53P} at $0.6< z < 0.9$. However, \citet{2014ApJ...782...33L} did not find evidence (out to \textit{z} $\sim$ 0.8) for an environmental dependence of the MS, although they did find a significant reduction of the sSFR by 17 per cent in cluster environments $(M_{\rm halo} >  10^{14} \text{M}_\odot)$. At z $\sim$ 0.5, \cite{2015ApJ...814...84D} also showed that [OII] EW (a measure of sSFR) versus stellar mass relation is independent of environment (filament vs. field), indicating the environmental invariance of the MS. Furthermore \cite{2014ApJ...796...51D} showed the environmental (filament, cluster, field) independence of the MS in the COSMOS field  \citep{2007ApJS..172..150S} at z $\sim$ 0.84.

At higher redshift ($z \sim 1.5$) \cite{2014ApJ...789...18K} found no direct evidence for an environmental dependence of the MS for H$\alpha$ emitters. Furthermore, \cite{2013MNRAS.434..423K} found that the difference between the field and cluster MS is less than 0.2 dex in redshifts smaller than $\sim 2$ based on H$\alpha$ emitters.

The dependence of the MS on the large scale environment is still under discussion, however a clear correlation seems to exist between SFR and paired galaxies. \cite{2015MNRAS.451.1482M} found that paired massive $(\log (M_*/M_\odot) > 11.5)$ galaxies have higher SFR than isolated galaxies, and this same result was found for lower mass galaxies with the galaxy pairs in the Sloan Digital Sky Survey \citep[e.g.][]{2010MNRAS.407.1514E,2012MNRAS.426..549S} and for major-merger pairs in \textit{Herschel} \citep{2015arXiv151206467C}.

No consensus has yet been reached on how the MS depends on environment or redshift.  This is partially because the aforementioned methods differ in how they select the galaxies, how they estimate star formation, and how they measure environment.  The selection of which method to use to determine the environment can cause differences in which galaxies are selected to be in a certain density regime, and therefore different results \citep{2012MNRAS.419.2670M}.

To accurately probe the normalisation of the MS, it is necessary to measure the SFR in a wide range of different environments over cosmic time. For the SFR we use the FIR data from the Herschel Multi-tiered Extragalactic Survey \citep[HerMES;][]{2012MNRAS.424.1614O}. In order to probe the environment and the stellar mass, we exploit the rich multi-wavelength data and volume of the COSMOS field. 

The Herschel Extra-galactic Legacy Project (HELP, \cite{2015arXiv150806444V}, Oliver et al. 2016, in prep.) aims to collate and homogenize observations from many astronomical observatories to provide an integrated data set covering a wide range of wavelengths from the radio to the UV. The key focus of the HELP project is the data from the extra-galactic surveys from ESA's \textit{Herschel} mission \citep{2010A&A...518L...1P}, covering over 1300 deg$^2$. HELP will add value to these data in various ways, including providing selection functions and estimates of key physical parameters. The data set will enable users to probe the evolution of galaxies across cosmic time and is intended to be easily accessible for the astronomical community. The aim is to provide a census of the galaxy population in the distant Universe, along with their distribution throughout the 3-dimensional space. 

Another key feature of HELP will be the generation of galaxy density maps. In this paper, we apply our chosen methodology for measuring density fields to publicly available data in the COSMOS field to explore the environmental dependence of star formation as probed by \textit{Herschel}.

The format of this paper is as follows. We describe the data we use in Section \ref{sec:data}. We describe our methods of determining the environment of the galaxy (Section \ref{sec:density}), our stacking analysis (Section \ref{sec:simstack}) and how we obtained SFRs (Section \ref{sec:SFR}). The results are described in Section \ref{sec:results}. The discussion and conclusions can be found in Sections \ref{discussion} and \ref{sec:conclusion}. We use a standard flat cosmology with $\Omega_M = 0.3$ and $H_0 = 70\ \text{km s}^{-1} \text{Mpc}^{-1}$.

\section{Data} \label{sec:data}

\subsection{The HerMES survey}
We use the SPIRE data \citep{2010A&A...518L...3G}  from the HerMES \citep{2012MNRAS.424.1614O}  survey to compute the SFRs of our sample of galaxies. We use all $250\,\mu$m, $350\,\mu$m and $500\,\mu$m SPIRE bands in the COSMOS field from the second data release, DR2 \cite[5$\sigma$  depth of 15.9, 13.3, 19.1 mJy; at 250, 350 and 500$\,\mu$m,  respectively,][]{2015ApJ...809L..22V}. The HerMES (and in future also the HELP) data can be obtained from the HeDAM database\footnote{\url{hedam.lam.fr}}.

One of the challenges at the longer wavelengths probed by SPIRE is extragalactic confusion \citep[e.g.][]{2010A&A...518L...5N}, whereby many sources detectable with higher resolution shorter wavelength imaging are located within a single SPIRE beam, and therefore appear as one SPIRE source. The SPIRE FWHM for 250, 350 and 500$\,\mu$m is 18.1, 25.5, and 36.6 arcsec, respectively \citep{2010A&A...518L...3G,2015ApJ...809L..22V}. To estimate the SPIRE flux density for individual galaxies we need to exploit prior information of the position, mass and redshift of the galaxies. We use a stacking method to obtain these flux density estimates, with a method that will be addressed in Section~\ref{sec:simstack}. 

\subsection{Multi wavelength catalogue}
Photometric redshifts for our sample are obtained from the COSMOS UltraVISTA  $K_{\rm s}$-band selected catalogue \citep{2013A&A...556A..55I, 2012A&A...544A.156M}. The catalogue contains 30 bands ranging from UV to NIR in broad, intermediate and narrow bands,  and contains 220 000 galaxies.  The photometric redshifts were obtained using the \texttt{Le\_Phare} code \citep{2006A&A...457..841I} and calibrated against spectroscopic redshifts. Due to the large range in wavelength, the availability of intermediate bandwidth photometric filters and the good quality of the data, the estimated redshifts are very accurate. For $z < 1.5$, \cite{2013A&A...556A..55I} obtained a precision of $\sigma_{\Delta z/(1+z)} = 0.008$ at $i^+ < 22.5$ ($<1\%$ catastrophic outliers) and even for faint ($i^+ \simeq 24$) sources the accuracy is better than 3\%. At higher redshift ($1.5<z<4$ ) the precision is given by $\sigma_{\Delta z/(1+z)} \approx 0.03$ \citep{2013A&A...556A..55I}. Furthermore the photometric redshifts are characterised by their full probability distribution function (PDF).

Stellar masses are derived from the SED using the Stellar Population Synthesis model of \cite{2003MNRAS.344.1000B} in combination with the \cite{2003PASP..115..763C} initial mass function. The stellar masses are model dependent and can vary by 0.1-0.15 dex depending on the model. The sample of star-forming galaxies is complete above a stellar mass of $\sim10^{8}M_\odot$ at $z = 0.2$ and or $ \sim 10^{10}M_\odot$ at $z = 3.0$.

\begin{figure}
  \centering  
  \includegraphics[trim = 00mm 00mm 00mm 00mm,width = 1.00\columnwidth]{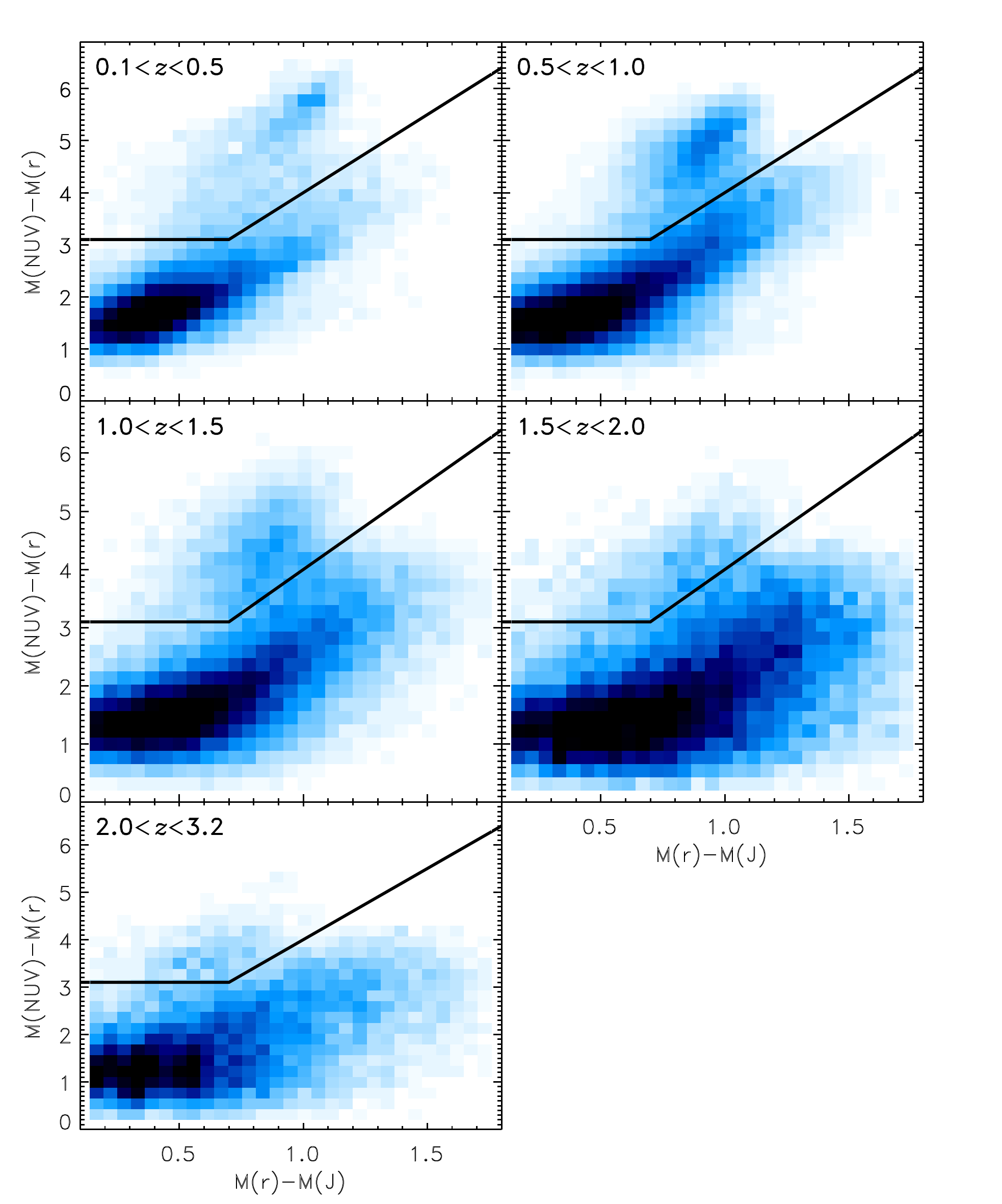}
  \caption{The colour selection used to separate the star-forming and quiescent galaxies. Galaxies with $M_{\rm{NUV}} -M_r > 3(M_r-M_J)+1$ and $M_{\rm{NUV}} - M_r > 3.1$ are considered to be quiescent.}
  \label{fig:colour}
\end{figure}

We treat the  star-forming and the quiescent galaxies differently, as we expect that their FIR properties to be quite distinct.  To discriminate we use the indicator from \cite{2013A&A...556A..55I}. This indicator is based on a rest-frame colour selection: galaxies with $M_{\rm{NUV}} -M_r > 3(M_r-M_J)+1$ and $M_{\rm{NUV}} - M_r > 3.1$ are considered to be quiescent (Figure \ref{fig:colour}).  This colour selection was chosen instead of a U-V vs. V-J selection because of the larger dynamical range. The NUV rest frame can also be sampled by optical data while the U band falls out of this wavelength range at $z>2$. Furthermore, the NUV-r seems to be a better indicator of current SF activity \citep[e.g.][]{2007ApJS..173..342M,2013A&A...556A..55I}. 

Because we only use the colour-colour selection we do not segregate starburst galaxies (galaxies that lie above the MS). This means that our average SFR estimates for the MS will be enhanced by 12.1 per cent relative to other methods which exclude the starburst galaxies from their sample \citep{2012ApJ...747L..31S}.  This 12 per cent represents the increase in the mean SFR changing from a single log-normal distribution for the MS only to a MS+starburst distribution as described by two offset log-normals \citep{2012ApJ...747L..31S}.

\section{Method}

\subsection{Sample selection}
Many methods have previously been used to probe the environment of galaxies. The most reliable methods of determining if a galaxy is located in a cluster, has close companions or resides in a dense environment rely on the use of spectroscopic redshifts. However, spectroscopy is time-consuming to obtain and is not practical for large numbers of galaxies over a large luminosity and redshift range, which are needed to exploit the full potential of HELP. 

To avoid this problem, we use photometric redshifts. The main disadvantage of photometric redshifts is that they are not accurate enough to associate a given galaxy with a given structure; the physical scale associated with the uncertainty in photometric redshift is normally much larger than the size of a galaxy cluster. However, for a large galaxy sample we can statistically infer that galaxies found in dense regions according to their photometric redshifts will also be in dense environments in real space \citep{2015arXiv150101398L}. 

Several methods have been developed to extract the environmental density of galaxies using their spatial distribution. Some of the most commonly used methods are the \textit{N}th nearest neighbour method (\textit{N} divided by the area containing \textit{N} neighbours), galaxy counts in a circular (adaptive) kernel, and the Voronoi tessellation method \citep[e.g.][]{2012MNRAS.419.2670M,2013ApJS..206....3S}.

The redshift range we used to make maps of the density of galaxies was selected carefully to optimise the accuracy of the density map. For lower redshifts the volume of the COSMOS field is too small to be useful, and does not accurately probe a range of environments. On the other hand, at higher redshift the photometric redshifts become more uncertain and the number densities decrease, so we restrict ourselves to the range $0.1<z<3.2$ \citep{2015ApJ...805..121D}. The typical photometric redshift error increases with magnitude, so we only consider galaxies with $K_{\rm{s}{\rm AB}} < 24$. We made this magnitude cut to use all available galaxies with accurate redshifts. However, by making this cut we only select relatively bright galaxies in the mass range for which we are incomplete. This can result in an overestimation of the mean SFR for low mass galaxies because we do not detect galaxies with a low SFR.  Note that we cannot see this effect in Figure \ref{fig:flux_vs_mass} because the galaxies are weighted according to their mass, leading to a very small contribution of the few galaxies below the mass limit. Furthermore, we only consider those galaxies outside the optically masked areas defined by \cite{2013A&A...556A..55I}.

To obtain sufficiently large samples, while exploring the evolution across time and environment, we divide galaxies into bins of redshift and density. We defined nine bins in environment and five in redshift so that each subset would contain approximately 11 per cent of the actively star-forming galaxies at that redshift. This yields $>1400$ active star-forming galaxies in every bin (Table \ref{tab:list}). 

In Section \ref{sec:density} we describe how we obtained the environmental information for our sample of galaxies, and in Section \ref{sec:simstack} we describe our method to assign flux densities to the galaxies with the use of stacking.

\subsection{Density estimates} \label{sec:density}

The density maps are constructed using the adaptive Gaussian kernel procedure from \cite{2015ApJ...805..121D}. This method uses a Gaussian kernel (with an adaptive width) to smooth the map, and therefore gives an estimate for the density at the scale of the kernel width. This choice of method was made on consideration of tables 3 and 4 of \cite{2015ApJ...805..121D}, where the kernel method performed best in simulations. 
Another advantage of the kernel method is its simplicity and the intuitive way in which the weights are assigned to a galaxy. We adopt the same adaptive kernel size, angular position cut, magnitude selection and overlap between redshift slices as used by \cite{2015ApJ...805..121D}. However, we make some changes in the application of weights and the edge corrections.

\begin{figure}
  \centering  
  \includegraphics[trim = 30mm 00mm 00mm 00mm,width = 1.05\columnwidth]{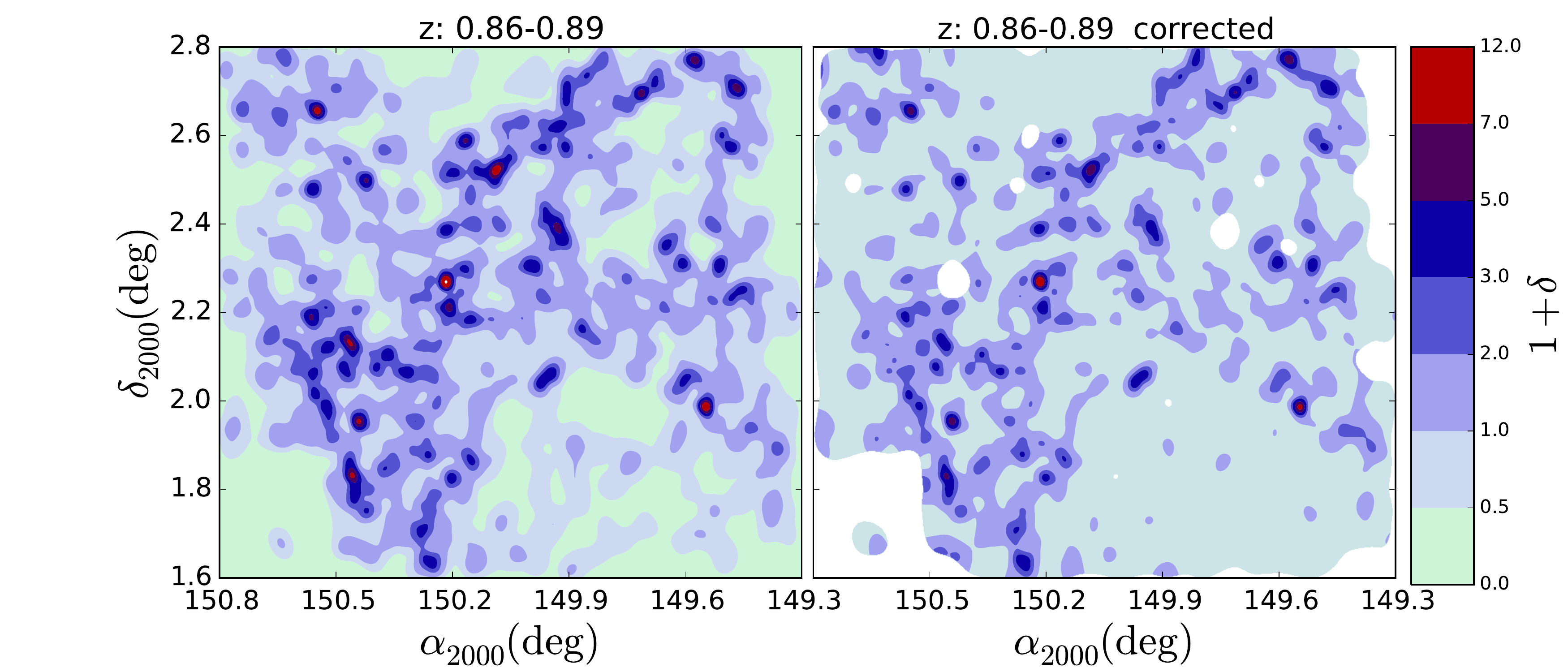}
  \caption{One of the redshift slices in the COSMOS field. On the left we show the density map, using essentially the method of \citet{2015ApJ...805..121D}, with slight modifications of the original method. On the right we show the same map but divided by the average of 40 mock maps. Regions where the mock map has a density less that half of the mean density in the slice are not taken into account (white areas).}
  \label{fig:map}
\end{figure}

Our method is as follows:

\begin{itemize}

\item We construct a series of redshift slices starting at $z=0.1$ and with a width $(\delta z = 2 \Delta z_{\rm med})$, where $\Delta z_{\rm med}$ is the median of the photo-z uncertainty of galaxies within that redshift slice. Each redshift slice starts in the middle of the previous slice. For galaxies without a second peak in the PDF (with a probability bigger than 5 per cent for the second peak) we make a Gaussian assumption for the shape of the PDF \citep{2015ApJ...805..121D}. 

\item Every galaxy is distributed between all slices according to the PDF \citep{2013MNRAS.433..771B}.  If a galaxy has a probability of 60 per cent to be in slice $a$ and 20 per cent to be in slice $b$ then the weight ($w$) in slice $a$ will be 0.6 and for slice $b$ will be 0.2. In \cite{2015ApJ...805..121D}, a galaxy can influence the density maps in adjacent slices since slices are overlapped.

\item Within a slice the local density ($\hat{\Sigma}_i$) at a galaxy position ($\bar{r}_i$) is determined by a weighted adaptive kernel estimator with a global width \textit{h} of 0.5 Mpc, following \cite{2015ApJ...805..121D}: 
\begin{align}
  \hat{\Sigma}_i &=  \frac{1}{\sum_{j= 1,j \neq i}^N w_j} \sum_{j = 1, j \neq i}^N w_i K(\bar{r}_i,\bar{r}_j,h), \\
  K(\bar{r}_i, \bar{r}_j, h) &= \frac{1}{2\pi h^2} \exp\left(  \frac{- |\bar{r}_i - \bar{r}_j |^2}{2h^2}\right),
\end{align}
where $\bar{r}_j$ is the position of a galaxy with weight $w_j$. Rather than adopting a uniform value for \textit{h} over the whole field, the local kernel width changes adaptively in accordance with the density of galaxies, with smaller kernel values in more crowded regions: 

\begin{equation}
  h_i = 0.5 {\rm Mpc}  \left( G/\hat{\Sigma} (\bar{r}_i) \right)^{0.5},
\end{equation}
where $\hat{\Sigma} (\bar{r}_i)$ is the galaxy density at position $\bar{r}_i$ calculated with \textit{h} = 0.5 Mpc and \textit{G} is the geometric mean of all $\hat{\Sigma} (\bar{r}_i)$. The density field ($\Sigma (\bar{r})$) is then obtained by

\begin{equation}
  \Sigma (\bar{r}) = \frac{1}{\sum_{i = 1}^N w_i} \sum_{i = 1}^N w_i K(\bar{r},\bar{r}_i,h_i),
\end{equation}
with $\bar{r} = (x,y)$ is the location in our 2D grid map.

\item As a convenient, dimensionless, measure of the galaxy environment, we define the over-density for a galaxy at position $\bar{r}$ by the density at that position in the map divided by the median density of every position in the slice:
\begin{equation}
  1+\delta = \frac{\Sigma(\bar{r})}{{\rm median}\left(\Sigma(\bar{r})\right)}.
\end{equation}

\item For scientific analysis, \cite{2015ApJ...805..121D} discarded those galaxies that were close to the edge or masked areas. We correct for the underestimation of densities near edges and masked areas using a different method. We create 40 mock maps in which the galaxies within a given redshift slice are given angular coordinates of galaxies selected randomly from all redshift slices. We divide the observed density field by the average of the mock density field. To avoid errors introduced by large corrections in the proximity of heavily masked regions, we exclude all areas in which the density in the mock map is less than half the mean. With this method we can still use galaxies relatively near the edge without introducing spurious low-density environments (see Figure \ref{fig:map}).

\end{itemize}

Our density maps optimally exploit the redshift PDF information for a smoothing kernel that is adaptively smoothed in the transverse direction, but is convolved with a discrete, top-hat kernel in the radial direction. In future work for HELP we will amend the method to provide an adaptive kernel in 3D. 

Having determined the density field, we can then assign a density to each galaxy. This assigned density is the measurement of the density at the angular position of the galaxy in the redshift slice where the photo-$z$ PDF is highest.

Since the absolute density and the over-density field evolve significantly with time through gravitational instability, we define the environment with reference to the surface density percentiles. In each redshift bin, we compute the density percentiles using every redshift slice within that redshift bin. We use these percentiles to create nine density bins, and we assign galaxies to the density bin appropriate to their density (Table \ref{tab:list}). To some extent, the environments defined by density percentiles are fixed with cosmic time, i.e. galaxies in the densest 5 per cent of the Universe today are expected to have been in the densest 5 per cent regions at an earlier time.

\begin{table*}
  \begin{tabular}{ | c | c c c c c }
    Density percentile & \multicolumn{5}{c}{Redshift ranges} \\
     &   0.1$<$z$<$0.5  & 0.5$<$z$<$1.0 & 1.0$<$z$<$1.5 & 1.5$<$z$<$2.0 & 2.0$<$z$<$3.2 \\ 
    & $N_{gal}$, $f_{sf}$&$N_{gal}$, $f_{sf}$&$N_{gal}$, $f_{sf}$&$N_{gal}$, $f_{sf}$&$N_{gal}$, $f_{sf}$\\ \hline
    0 -- 40       & 4148  \ \  0.91 & 8353  \ \  0.91& 7505  \ \  0.90& 5756  \ \   0.95& 3493  \ \ 0.95\\
    40 -- 55     & 3037  \ \  0.91 & 6471  \ \  0.90& 5612  \ \  0.90& 3528  \ \   0.94& 2870  \ \ 0.95\\ 
    55 -- 65     & 2751  \ \  0.92 & 5480  \ \  0.89& 4739  \ \  0.90& 2869  \ \   0.94& 2384  \ \ 0.95\\
    65 -- 75     & 3292  \ \  0.89 & 6981  \ \  0.88& 5666  \ \  0.90& 3368  \ \   0.95& 2952  \ \ 0.95\\ 
    75 -- 85     & 4292  \ \  0.89 & 9247  \ \  0.87& 7086  \ \  0.90& 3995  \ \   0.95& 3607  \ \ 0.95\\
    85 -- 90     & 2770  \ \  0.86 & 5694  \ \  0.86& 4342  \ \  0.89& 2358  \ \   0.94& 2213  \ \ 0.95\\
    90 -- 95     & 3587  \ \  0.85 & 7274  \ \  0.84& 5040  \ \  0.89& 2688  \ \   0.94& 2509  \ \ 0.96\\
    95 -- 97.5  & 2401  \ \  0.82 & 4581  \ \  0.83& 3057  \ \  0.89& 1614  \ \   0.93& 1489  \ \ 0.96\\
    97.5 -- 100& 4048  \ \  0.66 & 7168  \ \  0.76& 3952  \ \  0.87& 2051  \ \   0.93& 1726  \ \ 0.97\\ \hline \hline
    all             &30326   \   0.85 &61249   \  0.86&46999    \  0.90&28277   \    0.94&23243 \  0.95 
  \end{tabular}
  \caption{Number of galaxies ($N_{gal}$) in the percentile bins we use for stacking, and the star-forming fraction ($f_{sf}$) Every bin of star-forming galaxies contains over 1400 galaxies, leading to a reliable stacked signal. The density percentile bins where chosen to approximately obtain the same number of galaxy in every density percentile bin, but for comparison we fixed the density percentile bins over redshift. Due to this combination there is a slight variation in the number of galaxies per density bin. }
  \label{tab:list}
\end{table*}

\subsection{SIMSTACK} \label{sec:simstack}

Our aim is to measure the average star formation activity of galaxies aggregated by redshift and environment, while taking into account variations across bins, e.g. in the empirical relation between star formation and stellar mass -- the ``main-sequence".  To do this we use a flux stacking technique, with a weighting scheme to account for these known variations. We are using a stacking technique to get around the confusion problem: for one individual galaxy we cannot say what the contribution from non-correlated background sources is, but the mean contribution for a random stacked sample goes to zero in a mean subtracted map.

We use \texttt{SIMSTACK} \citep{2013ApJ...779...32V} as our stacking tool. \texttt{SIMSTACK} simultaneously estimates the average flux density for a number of samples of galaxies, modelling the SPIRE map by assuming that all galaxies in this sample to have the same flux density. \cite{2013ApJ...779...32V} segregated galaxies according to their stellar mass and  redshift,  and characterised how the FIR emission depended on these parameters. 

The \texttt{SIMSTACK} algorithm has been used and tested by several papers \citep{2014MNRAS.437..437A,2015MNRAS.454..419B,2015A&A...573A.113B,2015ApJ...809L..22V,2015ApJ...809..173W,2016ApJ...816...48N}. It works optimally for large samples of galaxies which are expected to have a similar flux density.

 Even in our highest redshift bin, we have over 23 000 galaxies simultaneously fitted by \texttt{SIMSTACK}, so random foreground and background sources will not affect our results.
 
In Table \ref{tab:list}, we list the number of galaxies in each redshift bin used in our stack. We ran \texttt{SIMSTACK} simultaneously on the star-forming and quiescent sample to avoid overestimating the SFR in dense environments due to confusion with nearby quiescent sources. Because \texttt{SIMSTACK} simultaneously fits all galaxies, it will give reliable values for the stack in both the field and for cluster galaxies. Only galaxies below the detection limit (in $K_{\rm s}$) and correlated with our target sample can affect the result. This effect should be larger in the clusters, but we expect these ``non detected galaxies'' to have low SFR (and low flux density in the SPIRE bands) and therefore they should not change our results very much. If they have any effect it would be to increase our estimates of the SFR in dense environments.

To account for the known internal variation within the bins we model the relationship between the FIR emission, stellar mass and the redshift of the galaxies. Since we are interested in relative measurements of SFR (in different environments) we do this by using weights in the stacking code. Essentially, these weights scale the contribution of each galaxy in the flux stack to what would be emitted by a reference galaxy at the centre of the redshift bin and with a reference stellar mass, see Section \ref{sec:red_w} and\ref{sec:mass_w} for more detail.

\subsubsection{Redshift weighting} \label{sec:red_w}

Within each redshift bin, there is a distribution of redshifts. The nearby ones will appear to be brighter, without having intrinsically higher luminosity or SFR. We correct for this effect following \cite{2010MNRAS.405.2279O}.  We weight the galaxies by $w_d$, which comes from the square of the luminosity distance $(D_L)$  relative to that of the middle of the redshift bin $(z_{\rm ref})$:

\begin{equation}
  w_d = \left( \frac{D_L(z_{\rm ref})}{D_L(z)} \right)^2.
\end{equation} 

Another adjustment originates from the \textit{K} correction, the SPIRE flux densities sample different parts of the rest-frame spectrum $(I_\lambda)$ for galaxies at different redshifts. We estimate the weight ($w_k$)  for the \textit{K} correction for an observed frequency ($\nu_0$)  and luminosity ($L_\nu$) to be the ratio of the rest-frame flux density for an object at redshift \textit{z} to that of the value at the middle of the bin:  
\begin{equation}
  w_k = \frac{1+z}{1+z_{\rm ref}} \frac{L_\nu ([1+z]\nu_0)}{L_\nu ([1+z_{\rm ref}]\nu_0)}.
\end{equation}

We use template SEDs provided by \cite{2013A&A...551A.100B}, which fits the median SED in the FIR (with at least seven FIR bands) for different spectral types of galaxies.  As a first approximation, we use the spiral galaxy template for the star-forming systems, and the elliptical template for the quiescent galaxies. Later this formed the basis for an iteration described at the end of Section~\ref{sec:mass_w}. 

Another weight $(w_e)$ arises from the known evolution in the MS with roughly $ (1+z)^\gamma$; that is if a galaxy has a higher redshift we expect a higher flux density due to relatively higher star formation:
\begin{equation}
  w_e = \left( \frac{1+z}{1+z_{\rm ref}} \right)^\gamma.
\end{equation}
Here we initially used $\gamma = 3$, and again this was the basis for an iteration. The overall effect of all these corrections is summarised in Table \ref{tab:correction}.  Since all of the corrections depend on redshift, we combine them to obtain a redshift dependent combined weight $(w_z)$ for every galaxy in the stack:

\begin{equation} \label{eq:w_z}
  w_z = w_d \times w_k \times w_e. 
\end{equation}

\begin{table}
  \begin{tabular}{llllll}    
     Redshift bin & \multicolumn{5}{c}{Weighting} \\    
      & \multicolumn{3}{c}{$\langle w_k^2\rangle$} &$\langle w_e^2\rangle$& $\langle w_d^2\rangle$\\ 
&      $250\,\mu$m & $350\,\mu$m & $500\,\mu$m &&\\\hline
    0.1 -- 0.5 &  0.040 & 0.067 & 0.099 & 0.038 & 3.552 \\
    0.5 -- 1.0 &  0.015 & 0.042 & 0.077 & 0.039 & 0.313\\ 
    1.0 -- 1.5 &  0.002 & 0.016 & 0.038 & 0.024 & 0.094\\ 
    1.5 -- 2.0 &  0.000 & 0.006 & 0.018 & 0.016 & 0.046\\ 
    2.0 -- 3.2 &  0.000 & 0.012 & 0.035 & 0.037 & 0.125\\
  \end{tabular}
  \caption{Mean square deviation of the redshift-dependent weights. Each weight, $w$, normalises galaxies at different redshifts to a reference point at the centre of the bin. Columns 2 to 4 show the weight that arises from the change in observed SED due to the \textit{K} correction. The fifth column shows the weight for the evolution of the MS over time. The last column shows the weight for the luminosity distance. For this table no distinction is made between galaxies in different environments.}
  \label{tab:correction}
\end{table}

\subsubsection{Mass weighting} \label{sec:mass_w}
To characterise the stellar mass dependence of the FIR emission, we follow the procedure explained by \cite{2013ApJ...779...32V}. We bin galaxies according to mass, redshift and galaxy type. We need several mass bins to obtain a good fit for the MS. For the star-forming galaxies we select the mass bins to contain either a total stellar mass of $10^{14} \rm{M}_\odot$ and a minimum of 100 galaxies, or $10^{15} \rm{M}_\odot$ and 50 galaxies. These mass bins where chosen so that each yield a clear detection of the stacked results in the SPIRE maps, and in the case that the slope of the MS is 1 these bins will all have approximately the same total signal. The quiescent galaxies are all placed in one bin.

From this set of stacked results, we can fit a mass vs. (redshift corrected) SPIRE flux density relation. This relation can be seen as a MS \citep{2011A&A...533A.119E} though with the redshift-corrected $(w_z$, Equation \ref{eq:w_z}) FIR flux density as a proxy for SFR. We exploit the fact that the integrated FIR flux density is expected to be proportional to the SFR, therefore we can use the model normally used to fit the MS: 
\begin{equation}   \label{eq:sfr}
   \log{\rm SFR} \propto \log{S_{\rm SPIRE}} = \alpha \log(M) +\beta. 
\end{equation}

Here $S_{\rm SPIRE}$ is the measured flux density with \texttt{SIMSTACK}.  We fit the parameters $\alpha$ and $\beta$, constraining the slope, $\alpha$, to be the same for the three bands. The results are shown in Figure \ref{fig:flux_vs_mass}. This enables us to apply a weight in comparison with a galaxy of reference stellar mass, $\rm{M}_{\rm ref}$:
\begin{equation} \label{eq:11}
  w_{\rm mass} = \frac{10^{\alpha \log(M) + \beta}}{10^{\alpha \log(M_{\rm ref}) + \beta}} = 10^{\alpha \log(M/M_{\rm ref})}.
\end{equation}
We set $M_{\rm ref}$ to a stellar mass of $10^{10} \rm{M}_\odot$, so that the stacked results give us the average flux density of a star-forming $10^{10} \rm{M}_\odot$ stellar mass galaxy at the middle of the redshift bin. We also use this slope $\alpha$ for the quiescent bin.

Having determined the weighting factors, we can use \texttt{SIMSTACK} to compute the mean, normalised SPIRE flux densities, aggregated in bins of redshift and environment. Our procedure also allows us to normalise the known variations with redshift and stellar mass to the centre of each redshift bin and for this reference stellar mass. 

Our results are independent of choice of $M_{\rm ref}$. If we had selected a different $M_{\rm ref}$ then $w_{\rm mass}$ would change for every galaxy accordingly, and the output of \texttt{SIMSTACK} would be the flux density of the new $M_{\rm ref}$ (we fit all the galaxies at the same time given the weights calculated in Equation \ref{eq:w_z},\ref{eq:11}). The underlying assumption for this is that all the galaxies follow the same slope in the Mass vs. SFR plane ($\alpha$, Equation \ref{eq:sfr}), at a certain redshift. With \texttt{SIMSTACK} we find the normalisation of this line at $M_{\rm ref}$ and it is this normalization that we track in different environments.

In total we had four runs with \texttt{SIMSTACK}. The first run was with the parameters described above, from which we got a first estimate for our best SED template, $\alpha$ and $\gamma$. For the second run we used these best values as input for our weights 
(Equation \ref{eq:w_z}, \ref{eq:sfr}). A third run was used to optimise the results for the fourth and final run, Section \ref{sec:SFR}.

\subsection{SFR estimation} \label{sec:SFR}

Having determined the mean, normalised SPIRE flux densities in each SPIRE band, we estimate a total integrated FIR luminosity (and hence SFR) for each redhift and density bin.  

We find the best fit SED through a least-squares fit from the library of \cite{2013A&A...551A.100B} to the mean normalised flux densities in the three SPIRE bands. The SPIRE bands probe the peak of the SED for intermediate redshifts and this gives us the most accurate SED normalisation. PACS data \citep{2010A&A...518L...2P} could be added and would probe the peak of the FIR emission better in our lowest redshift bin; however, due to differences in the map-making procedure (i.e. the non linear map making in PACS), and higher noise in the PACS data, we choose not to include these data in our analysis to avoid introducing biases into the sample \citep{2011A&A...532A..90L}.

Different SEDs are allowed for star-forming galaxies and passive galaxies and for different redshift slices. However, we use the same template for different stellar mass bins and environments at the same redshift. The best templates for every redshift bin are listed in Table \ref{tab:template}.

With this SED template, we compute the total  FIR luminosity $L_{\rm FIR}$ integrated over the rest frame spectrum ($L_\nu$) between 8$\,\mu$m and 1000$\,\mu$m.  This process is performed iteratively with the weighting processes in Section~\ref{sec:simstack}, i.e. applying the \textit{K}-correction using the optimum template. 

We then compute the SFR from $L_{\rm FIR}$, using the following calibration \citep{1997MNRAS.289..490R,2008MNRAS.386..697R,2010MNRAS.405.2279O}\footnote{This calibration is based on a \cite{1955ApJ...121..161S} IMF, to convert to other mass functions we refer to \cite{2014ARA&A..52..415M} and \cite{1997MNRAS.289..490R} for conversion factors.};

\begin{equation}
  \frac{L_{\rm FIR}}{\mbox{L}_\odot} = 0.51 \times 10^{10} \frac{\mbox{SFR}}{\mbox{M}_\odot \mbox{yr}^{-1}}.
\end{equation}
Here the fraction of ultraviolet energy absorbed by dust has been assumed to be $\epsilon = 2/3$. Because we are stacking, we use a fixed value of $\epsilon$, but ideally we would need this value for every galaxy. The HELP project will eventually assist in obtaining more information about the variation of $\epsilon$, but that is beyond the scope of this paper.

\section{Results} \label{sec:results}

\subsection{SFR in different environments} 

\begin{figure*}
  \centering  
  \includegraphics[trim = 5mm 00mm 00mm 00mm,width = 2.05\columnwidth]{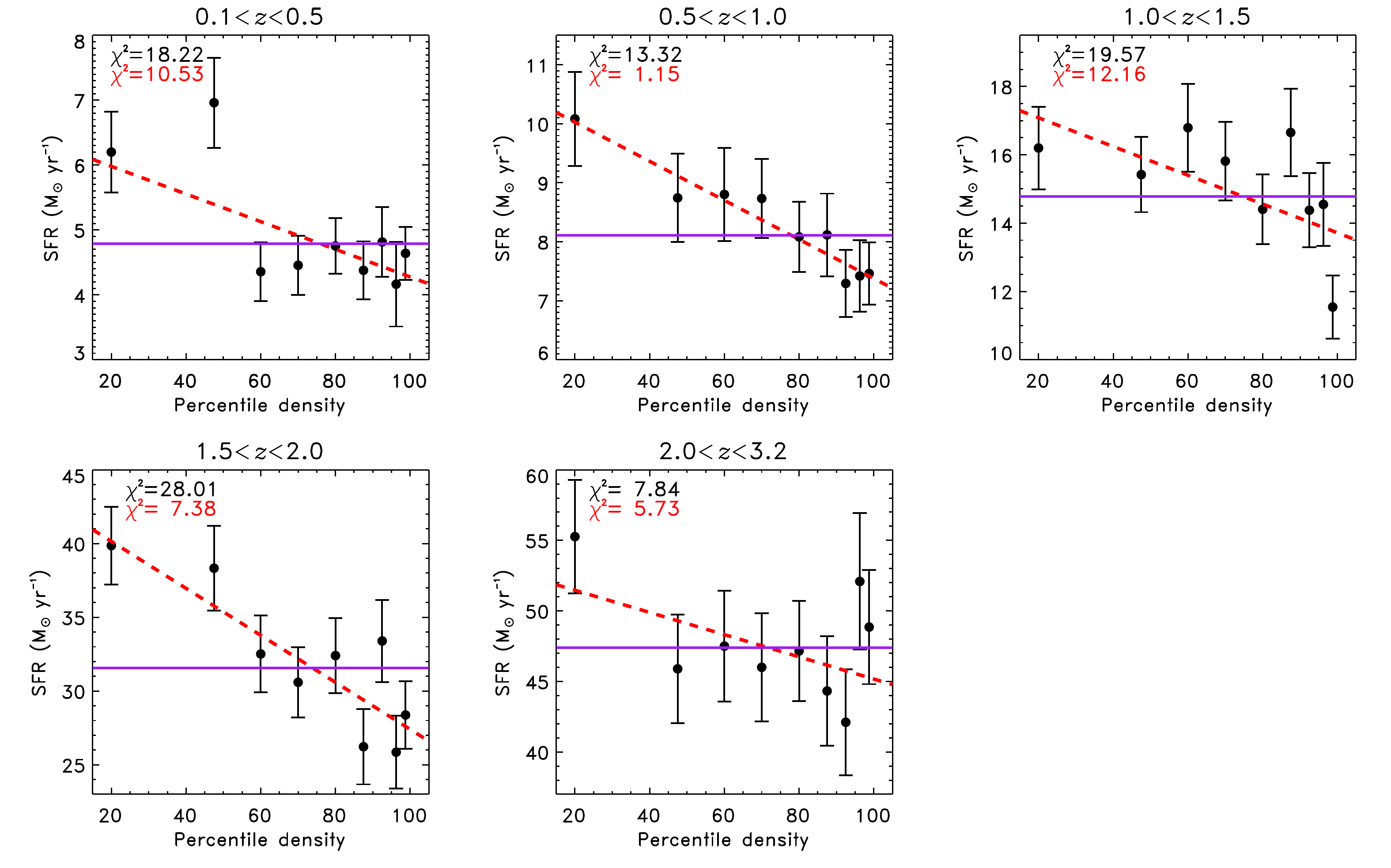}
  \caption{SFR of a $10^{10} \rm M_\odot$ stellar mass star forming galaxy in the COSMOS field vs. environment density for different redshift bins.  $10^{10} \rm M_\odot$ is the used $M_{\rm ref}$ in Equation \ref{eq:11}, and is the normalisation of the MS with slope $\alpha$ at the given density percentile. The black symbols represent the weighed mean of the calculated SFR of the three SPIRE bands (Figure \ref{fig:SFR_2}). The purple line represents the average value, the value of the SFR which should arise from a constant MS over different environments. In the top left corner, the $\chi^2$ value for an environmental-independent star formation (8 degrees of freedom, $N_{ \rm dof}$) is noted in black. It is clear that the star formation rate increases at higher redshift (the mean SFR is significantly higher in every higher redshift band). At intermediate redshift ($0.1<z<2$) a simple toy model (the line with the lowest $\chi^2$ in dashed red, 7 degrees of freedom) of a declining MS over environment seems to be a better fit (has a lower $\frac{\chi^2}{N_{ \rm dof})}$). In the redshift range $1.5<z<2$ this toy model has a reduced $\chi^2$ around one (indicating a good model), where the purple line has a $\frac{\chi^2}{N_{ \rm dof})} >$ 3 indicating that we can exclude this model for this redshift range. This effect is small (though marginally significant in a statistical sense), though small and both models (red and purple) are well within the intrinsic scatter of the SFRs in the MS. }
  \label{fig:SFR}
\end{figure*}

Our resulting SFR for the reference stellar mass in different environments and different redshifts are shown in Figure~\ref{fig:SFR}. We test our results against a constant MS for all environments and a toy model for which we fit a straight line to the SFR vs. percentile density. A fit to the evolution of this reference SFR (i.e. a normalisation of the main-sequence) yields $(1+z)^{2.4}$; this evolutionary rate was used to iteratively re-calculate the weights, $w_e$, used in Section \ref{sec:simstack}.

We construct the error bars $(\sigma_{tot})$ as a quadrature sum of the jackknife error ($\sigma_{jk}$), which covers the random errors associated with the sample variations within a bin) and the error $(\sigma_{z})$ from re-sampling our redshifts from the PDF (which covers the systematic errors from the uncertainty in the redshift of each galaxy):

\begin{equation}   \label{eq:err}
  \sigma_{tot}^2 = \sigma_{JK}^2 +  \sigma_{z}^2. 
\end{equation}

From Figure \ref{fig:SFR} we can see that there is no dramatic trend in the reference SFR as a function of environment at any epoch. Because our reference SFR has scaled every galaxy to the MS this indicates that the MS is roughly the same in every environment. However we can confirm that the SFR for star-forming system increases over cosmic time with roughly $(1+z)^{2.4}$. 

An additional subtle trend is worth noticing. In the range $0.1 < z < 2$ there appears to be a slight decline in SFR towards higher densities. We quantified this by calculating the reduced $\chi^2$ for a declining toy model. This toy model (red dashed line in Figure \ref{fig:SFR}) has a lower reduced $\chi^2$ in all of our redshift bins, indicating a lower MS in dense environments. This is in agreement with \cite{2010ApJ...710L...1V} and \cite{2011ApJ...735...53P} who found a lower SFR for star-forming galaxies in cluster environments.  Similarly  \cite{2015ApJ...806....3A} found that the mean observed F814W$-$F160W colours for star-forming cluster galaxies at $z \sim 2.1$ are 20 per cent ($3.6\, \sigma$) redder (indicating a lower SFR) than for field galaxies at the same masses, indicating a suppressed MS. Note that in Figure \ref{fig:SFR} we have not included the systematic error on the SFR, because the effect of taking the wrong SED template is to move all data points up or down together. We have omitted this error for our comparisons of different regions (see Appendix \ref{app:errors} for more detail about the error analysis).

In the redshift range $1.5 < z < 2$ the declining toy model has a reduced $\chi^2$ close to one, indicating a good fit, where the environment independent model has a reduced $\chi^2$ greater than three. We conclude that our data at $1.5 < z < 2$ is inconsistent with the hypothesis of an environmental independent MS with significance at level of 1 per cent, measured using \textit{p}-value. However, this is a small effect and all our data falls well within the 0.2 dex intrinsic scatter of the SFRs in the MS. In the other redshift bins we cannot exclude the simple hypothesis. 

We can also use the stacked, normalised, SFRs to assign an estimated SFR to every galaxy  (taking into account its stellar mass and the weights applied).  With this SFR for each galaxy we can produce estimates for the SFR-density in the COSMOS field. 

\subsection{Cosmic variance}

Figure \ref{fig:SFR} does not include the effect of ``cosmic variance", i.e., the possibility that our measurements in the COSMOS field, may not be representative of the Universe as a whole\footnote{We use the term ``cosmic variance" here, as is common in the galaxy cluster literature, although we appreciate the phrase is also used to refer to uncertainties due to the finite size of the observable Universe, and therefore some prefer ``sampling variance" for uncertainties for a finite field size.}.  We consider this to be an uncertainty only in the environmental metric, i.e. that the characterisation of the SFR for a population is unaffected by cosmic variance but that the density percentile ascribed to that population is.

Our primary environmental metric is the percentile of the density field. This is based on the over-density estimate, $\delta$, and so we consider the uncertainty in this in the following way:  

The fractional error in over-density, $\delta$, for dark-matter halos, or galaxy populations in a finite volume, is determined by the statistics of the density field and is the normal ``cosmic-variance" metric. The ``cosmic-variance" depends on the geometry of the field and the clustering strength or ``bias" of the population under consideration.  \cite{2011ApJ...731..113M} provide a tool, \texttt{getcv}, for calculating this variance using a halo occupation model to characterise the galaxy bias and clustering as a function of  redshift and stellar mass. Our populations are segregated by local environment rather than stellar mass, and so we cannot use this tool directly.  Instead we approximate it by assuming that bias follows the rarity of the samples under-consideration, e.g.  if we take the galaxies in the top 11 per cent of dense environments we assume that they will have the same bias (and thus cosmic variance) as the 11 per cent most massive galaxies (as in abundance matching).  Using the same stellar mass function as \cite{2013A&A...556A..55I} then allows us to estimate the stellar mass of galaxies with the same abundance.

To map this uncertainty in over-density to an uncertainty in percentile density is less straightforward.  We take a conservative approach, assuming that the uncertainty in percentile is compounded by the uncertainty in the mean density of the COSMOS field as a whole. Again we use the code \textit{getcv} to determine cosmic variance for dark matter halos in our redshift bin, which can be translated to the uncertainty on the mean density of the field. 

Combining the variance estimates on the density for certain type of galaxy in quadrature with the variance on the mean density provides us with an estimate of the effect of cosmic variance on our galaxy bins (see Table \ref{tab:cm_v}).
 
\begin{table}
  \begin{tabular}{lc}
   Redshift bin & Effect of cosmic variance in ($1+\delta$) \\ \hline
    0.1 -- 0.5 &  13.1--14.5 \% \\
    0.5 -- 1.0 &  7.2-- 8.1 \% \\ 
    1.0 -- 1.5 &  6.0--7.5 \% \\ 
    1.5 -- 2.0 &  5.9--7.8 \% \\
    2.0 -- 3.2 &  5.0--10.0 \% \\
  \end{tabular}
  \caption{Cosmic variance quantified in percentage error on $1+\delta$ over our redshift bins. The two values in the second column represent the cosmic variance for the lowest and highest density region respectively.}
  \label{tab:cm_v}
\end{table}

This effect could be represented as a horizontal error bar in Figure \ref{fig:SFR}.  However, it should be kept in mind that this is a systematic effect and not a measurement error. The values in Table \ref{tab:cm_v} suggest that if we had carried out the same analysis on a different part of the sky, we would have found different values of $1+\delta$ for the galaxies. If one wishes to compare our absolute results with those from a different field this effect should be taken into account. However, the  density estimates for our individual galaxies originate from the same field,  and so the relative environmental ranking should be unaffected by cosmic variance, for this reason we did not plot this error bar in Figure~\ref{fig:SFR}. 

An example of the effect of cosmic variance in the COSMOS field is the $z =0.73$ large scale structure found by \cite{2007ApJS..172..254G}. The effect of such a structure is that the mean density of that particular redshift slice increases. Therefore the overdensities assigned to the galaxies in that redshift slice will be slightly lower than the overdensity we assign to a galaxy in a similar environment in another redshift slice.

\subsection{SFR density}

\begin{figure}
  \centering  
  \includegraphics[trim = 5mm 00mm 00mm 00mm,width = 1.02\columnwidth]{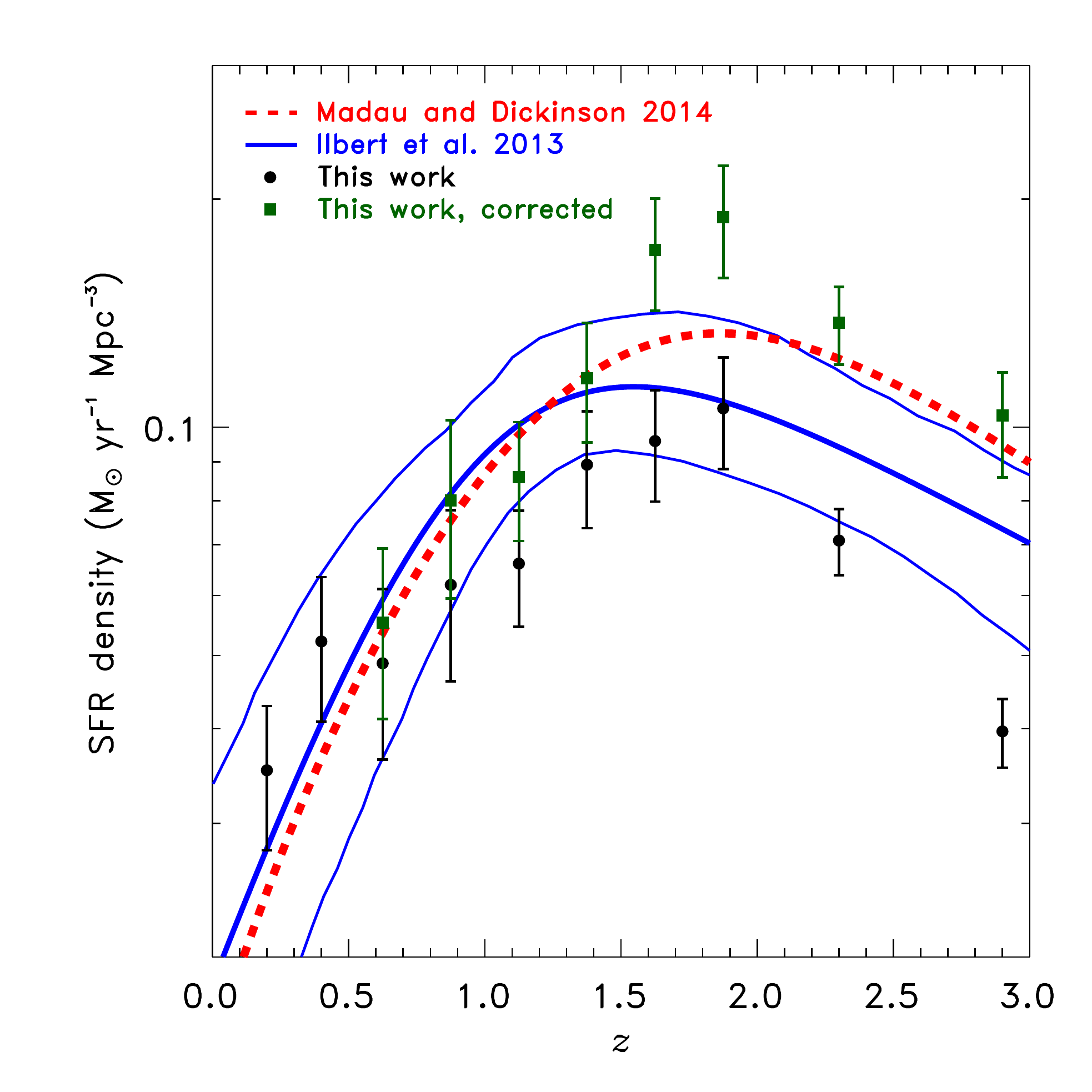}
  \caption{SFR density for galaxies with stellar mass $> 10^8 \rm M_\odot$ as function of redshift (black symbols); error bars include variance in the bias from choosing different templates, as well as the jackknife errors over the map. The green squares represent our completeness-corrected sample (and include the uncertainty associated with this correction).  For comparison the results from \citet{2014ARA&A..52..415M} and \citet{2013A&A...556A..55I} are shown in red (dotted) and blue respectively. The results from \citet{2013A&A...556A..55I}  are converted to a Salpeter IMF using the conversion constant from \citet{2014ARA&A..52..415M}.}
  \label{fig:density_2}
\end{figure}

With our estimates of individual galaxy SFRs we can calculate the SFR density (for galaxies with stellar mass $> 10^8 \rm{M}_\odot$) of the COSMOS field, and we plot this in Figure \ref{fig:density_2}. We correct for incompleteness by using the mass function of \cite{2013A&A...556A..55I} to calculate the number (and the mass) of galaxies which we do not observe. With our estimate for the MS we can assign a SFR to these galaxies and add this to the observed SFR density.

We estimate errors in these SFR densities using jackknife samples over the map combined in quadrature with errors from the mass function correction \citep{2013A&A...556A..55I} and an estimate of the systematic error in the template fitting.  The full error analysis is discussed in Appendix~\ref{app:errors}. 

Our SFRD results follow the curve of \citet{2014ARA&A..52..415M}, with only a difference in the peak which is higher by a factor of $1.4^{+0.3}_{-0.2}$. This result is in agreement with recent SFRD estimates from the FIR, using 500 $\mu$m detected sources \citep{2016arXiv160503937R}.  

\begin{figure}
  \centering  
  \includegraphics[trim = 5mm 00mm 00mm 00mm,width = 1.02\columnwidth]{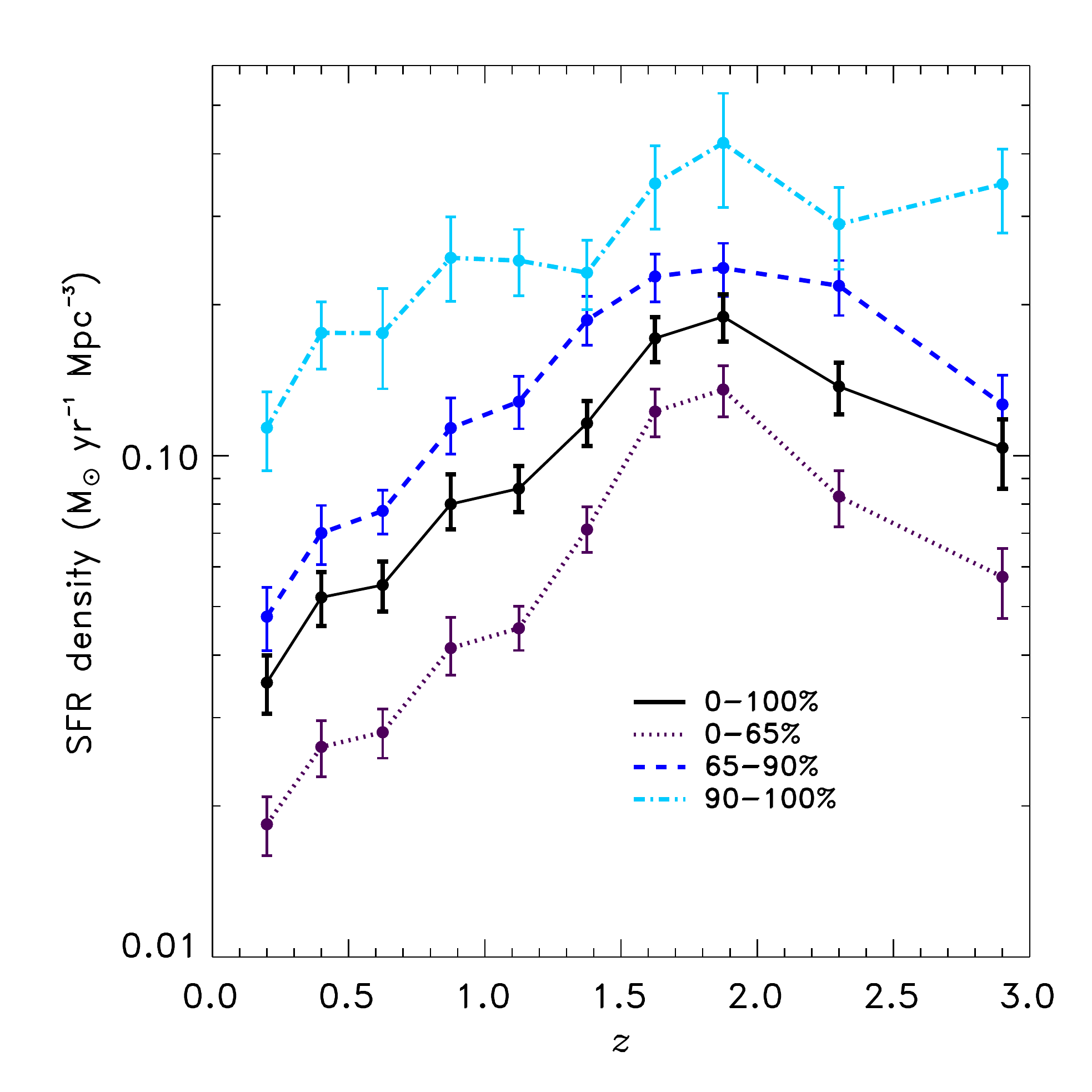}
  \caption{SFR density for four different percentile density regions as a function of redshift. The volume used to calculate the density is the volume of the percentile region, not the total volume of the field. There are no significant differences in the evolutionary trends of the four sub-samples. This figure also shows that although galaxies corresponding to the highest density sample themselves might have a lower star formation per unit mass, this population of galaxies still has a higher star formation per unit volume than the low density sample at every redshift. All data points have been corrected for incompleteness, but the bias in the templates is not taken into account here.} 
  \label{fig:density_all}
\end{figure}

We can also use the same SFRs and arrange the galaxies over the density regions to obtain SFR density estimates for different density environments over cosmic time (Figure \ref{fig:density_all}).  With this analysis, we cannot see a significant difference in the evolution of the SFR density for different environments. From our highest to lowest density sample we find a 73\%, 79\% and 86\% decline in the SFRD.

 This result is in slight tension with \cite{2015MNRAS.450.2749G} who found a steeper decline in the SFRD for cluster galaxies than for field galaxies.  \cite{2015MNRAS.450.2749G} used local cluster/field galaxies and probed the SFRD(z) by constructing the SFR-history of these samples. We look at the total instantaneous SFR as a function of cosmic time and environment. These two different ways to determine the SFR could lead to different results \citep[e.g.][]{2015MNRAS.451.2681S}.

\section{Discussion} \label{discussion}

It is interesting to explore possible explanations for the weak evidence that the typical SFR might be lower in denser environments, particularly in the range $1.5<z<2$. This lower MS was previously found at redshifts lower than 1 \citep[e.g.][]{2010ApJ...710L...1V,2011ApJ...735...53P,2016ApJ...816L..25P}.  This result may be spurious if the photometric redshift errors are significantly larger for the more extreme star-forming systems (i.e. those with higher sSFR) at higher redshift, scattering them out of overdense regions. But if real, this could be very interesting.  These redshifts correspond to the epoch of the peak of star formation activity and it is possible that we are actually witnessing the transition from star-forming to quiescent galaxies. At $ z>2$ the densest regions appear to follow the same star formation relation as lower densities.  However, below $z<2$ those star-forming galaxies might be expected to fall into the cluster and star formation begins to shut down. During the first part of this process these galaxies will still be classified as ``star-forming'', but the star formation rate is reduced relative to the stellar mass, lowering the average.  This explanation would fit with results that show that \textit{Herschel} maps exhibit strong clustering, compatible with halo models in which star formation at $z\sim2$ occurs in rich groups \citep{2013ApJ...772...77V}. 

It is important to remember that the subtle, but statistical significant differences we find in the MS over different environments is smaller than the intrinsic scatter (of 0.2 dex) of the SFR in the MS. Our results come from a mean stack which includes starburst galaxies, i.e. galaxies off the MS. These small environmental effects may come from MS galaxies or starburst galaxies. At low redshift starburst are merger driven and more prominent in intermediate and less dense
environments \citep[e.g.][]{2012MNRAS.426..549S,2014ARA&A..52..415M}. However at higher redshift the clustering of Herschel sources \citep{2010A&A...518L..22C} and maps \citep{2011Natur.470..510A,2013ApJ...772...77V} indicate that galaxies with high SFR are found in denser environments.

At $z < 1.5$ and $z>2.0$ we cannot formally exclude a hypothesis that the star formation rate verses stellar mass relation (i.e. the ``main-sequence") is the same for every environment.  That hypothesis has been supported by other observations,  although it is somewhat surprising theoretically, implying that environmental effects can change the relative proportion of galaxies that are star-forming or passive, but not the average SFR of the star-forming galaxies themselves \citep[e.g.][]{2010ApJ...721..193P}.  This implies that the environmental effects result in a rapid truncation of star formation \citep{2016arXiv160503182D}.

This tantalising result raises several questions, which are beyond the scope of this paper. For example, can we confirm this weak trend with better statistics including other fields with less cosmic variance, and where we can see a broader range of environments? Does our result depend on how we classify galaxies to  be star-forming or quiescent? How do our results depend on the accuracy of the  photometric redshifts, both within COSMOS and extending to regions with poorer phot-$z$ estimate? This shows the exciting opportunities that will come from exploiting the whole HELP data set, which will enable such analysis using multi-wavelength data over several fields.

\section{Conclusions} \label{sec:conclusion}

We have undertaken an investigation of the dust-obscured star formation activity as a function of environment and redshift.  We constructed a galaxy density field using an adaptive kernel smoothing and exploiting the full photometric redshift probability distribution function from the deep optical, NIR and IRAC data in the COSMOS field.  We characterised the density fields in terms of percentiles to facilitate comparisons between redshifts. We employed a ``stacking'' technique to estimate the normalisation of the ``main-sequence''  (i.e. the correlation between the SFR and stellar mass). This techniques  fits the Herschel SPIRE data from HerMES to all galaxies with photometric redshifts and stellar masses in the same redshift bin simultaniously.

A simple model in which the mean specific star formation rate for star-forming galaxies declines with increasing environmental density gives a better description at $0.1 < z < 2$ and is significantly better at $1.5<z<2.0$  with a reduced $\chi^2\sim1$ (q.v. $\chi^2\sim3$ for constant normalisation).  At $z < 1.5$ and $z>2.0$ we cannot exclude a simple hypothesis in which the main-sequence for actively star-forming systems is  independent of environment over the range.
We also estimate the evolution of the universally averaged star formation rate density in the COSMOS field and we find similarly strong evolution to previous studies though with a $1.4^{+0.3}_{-0.2}$ times higher peak value of the star formation rate density at $z \sim 1.9$.  When deconstructing the contributions to this evolution by density percentiles we do not see any significant differences in the shape of the evolution and note that the higher density regions of the Universe contribute more to the cosmic star formation history despite having a lower specific star formation rate.   

This works demonstrates the power of the \textit{Herschel} SPIRE data when coupled with high-resolution data sets and demonstrates  methodology that we will build upon to extend these studies to rarer higher density regions when exploiting the full 1300 deg$^2$ of data from the Herschel Extragalactic Legacy Project, HELP.

\section{acknowledgement}

We thank the referee for very useful comments that improved the quality of the work. This project has received funding from the European Union’s Horizon 2020 research and innovation programme under grant agreement No 607254. This publication reflects only the author’s view and the European Union is not responsible for any use that may be made of the information contained therein. Steven Duivenvoorden acknowledges support from the Science and Technology Facilities Council (grant number ST/M503836/1). Seb Oliver acknowledges support from the Science and Technology Facilities Council (grant number ST/L000652/1). Behnam Darvish acknowledges financial support from NASA through the Astrophysics Data Analysis Program (ADAP), grant number NNX12AE20G. E. Ibar acknowledges funding from CONICYT/FONDECYT postdoctoral project N$^\circ$:3130504. M Vaccari acknowledges support from the Square Kilometre Array South Africa project, the South African National Research Foundation and Department of Science and Technology (DST/CON 0134/2014) and the Italian Ministry for Foreign Affairs and International Cooperation (PGR GA ZA14GR02).

SPIRE has been developed by a consortium of institutes
led by Cardiff Univ. (UK) and including Univ. Lethbridge
(Canada); NAOC (China); CEA, LAM (France); IFSI, Univ.
Padua (Italy); IAC (Spain); Stockholm Observatory (Sweden);
Imperial College London, RAL, UCL-MSSL, UKATC,
Univ. Sussex (UK); Caltech, JPL, NHSC, Univ. Colorado
(USA). This development has been supported by national
funding agencies: CSA (Canada); NAOC (China); CEA,
CNES, CNRS (France); ASI (Italy); MCINN (Spain); SNSB
(Sweden); STFC, UKSA (UK); and NASA (USA).

\appendix

\section[]{Detailed information about the galaxies and templates used in the stack}

\subsection{The main-sequence fit}
\begin{figure}
  \centering  
  \includegraphics[trim = 5mm 00mm 00mm 00mm,width = 1.05\columnwidth]{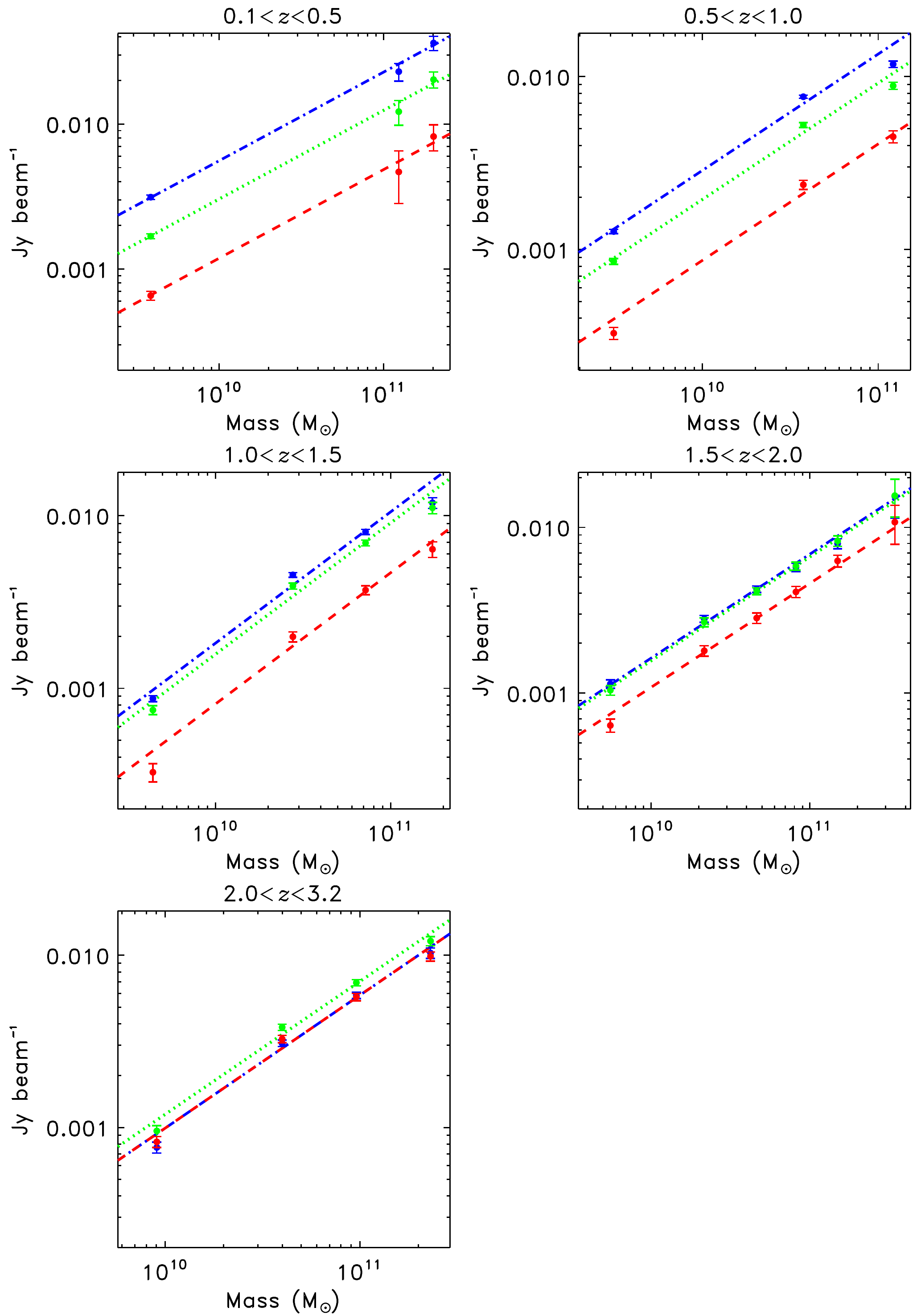}
  \caption{Stacked flux densities for the three different SPIRE bands, $250\,\mu$m (blue, dash-dot), $350\,\mu$m (green, dotted) and $500\,\mu$m (red, dashed) plotted against the stellar mass of the galaxies for different redshifts. The dashed lines are the best fits of the model (Equation \ref{eq:sfr}). For every redshift and every SPIRE band we can see a clear correlation of mass with flux density, so this plot is effectively another way to show the MS. }
  \label{fig:flux_vs_mass}
\end{figure}

To be able to probe the environmental dependence of the MS, we need to obtain the weights in the flux density contribution for every galaxy in this MS. For this purpose we used the fit from Figure \ref{fig:flux_vs_mass}.

This fit can be used to predict the flux density for a galaxy with a given mass, and so make an estimate of its weight in the stacking. In combination with the choice of best SED (as shown in Table \ref{tab:template}) we could apply the K-correction weights to determine a weight for every galaxy within the stack. By running \texttt{SIMSTACK} with these estimates we are able to investigate the offset of the MS in a certain region of the Universe.

\begin{table}
  \begin{tabular}{lll}
     Redshift bin & star-forming template & Quiescent template \\ \hline
    0.1 -- 0.5 &  Mod-SF-glx & Cold-glx \\
    0.5 -- 1.0 &  SF-glx-1 & Cold-glx \\ 
    1.0 -- 1.5 &  Ly-break & Cold-glx\\ 
    1.5 -- 2.0 &  WeakPAH-SF-flx-1 & Blue-SF-glx\\
    2.0 -- 3.2 &  Si-break & Spiral \\
  \end{tabular}
  \caption{SED templates used in the final run; for detailed information about the templates see  \citet{2013A&A...551A.100B}. }
  \label{tab:template}
\end{table}

\begin{figure*}
  \centering  
  \includegraphics[trim = 5mm 00mm 00mm 00mm,width = 2.05\columnwidth]{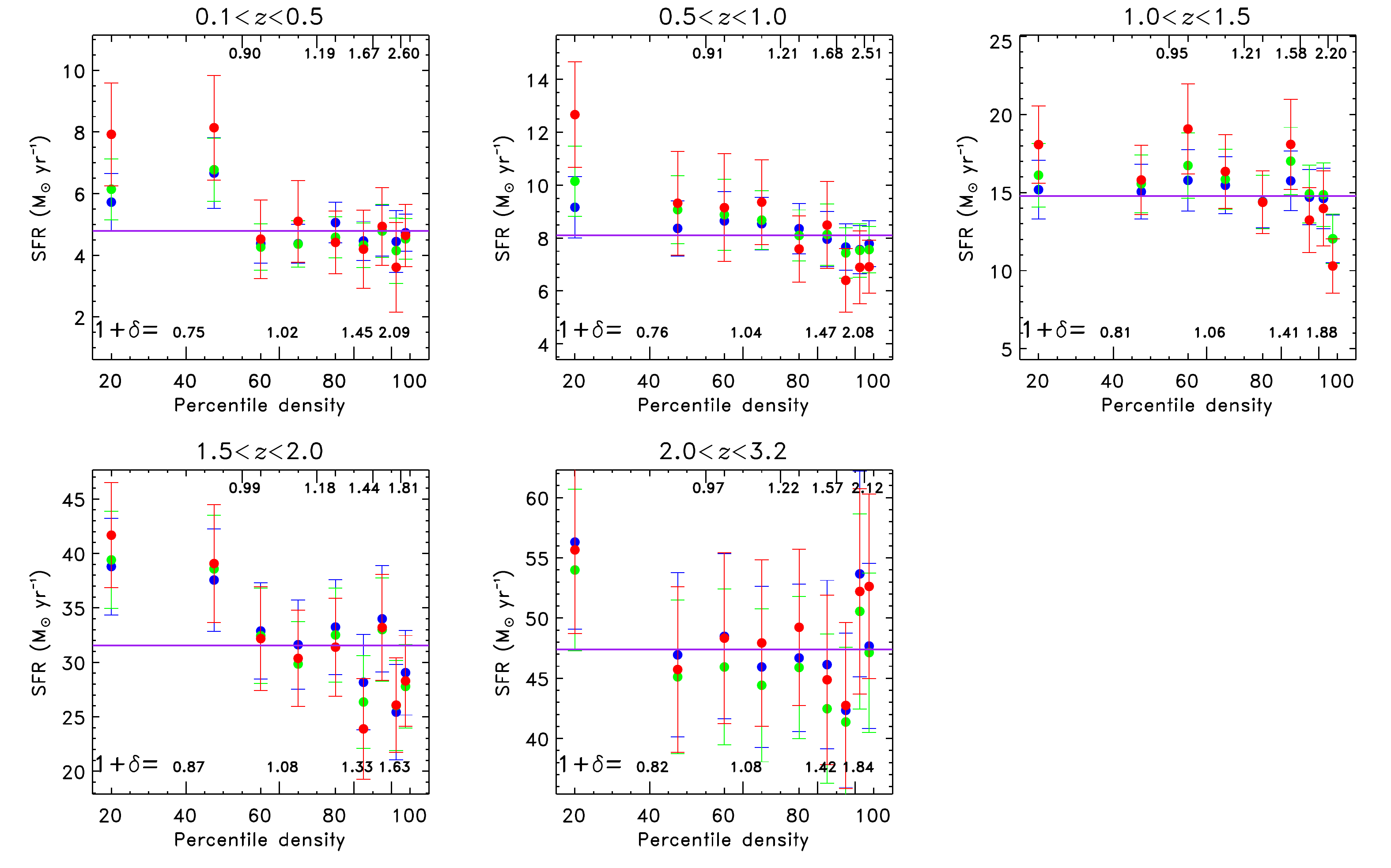}
  \caption{SFR from the stacked $K_{\rm s}$ selected star-forming galaxies in the COSMOS field for different redshifts and environments. The blue, red and green symbols represent the estimates from the $250\,\mu$m, $350\,\mu$m and $500\,\mu$m SPIRE bands. The purple line represents the average value, the value of the SFR which would arise from a constant MS over different environments. }
  \label{fig:SFR_2}
\end{figure*}

Our final result is the MS as function of environment, as measured by each single SPIRE band as shown in Figure \ref{fig:SFR_2}.  For every data point we constructed the jackknife errors over the map and the errors associated with a re-sampling of the data from the redshift PDF (Equation \ref{eq:err}). The estimates for the SFR from the three different SPIRE bands are in line with each other (have a reduced $\chi^2 \lesssim 1$ ), and we combine them to construct Figure \ref{fig:SFR}.

\section[]{Error estimation}\label{app:errors}

For Figure \ref{fig:SFR}  we constructed the errors by using both the variance over the map (using jackknife) and in redshift space. In this calculation we assumed that we had the correct SED to transform from the SPIRE flux density to SFR. 

If instead we had chosen an SED with an FIR peak with an offset from the intrinsic one, our three SPIRE estimates would have given very different SFR values, allowing us to rule out this FIR peak location. But if the peak is only slightly wrong, or if all three SPIRE bands are longward of the FIR peak, then several SED templates (with different SFRs) would all give a reasonable fit. On the other hand, the \textit{K}-correction, and the other corrections applied to obtain to our stacking list would still be roughly the same for these ``good'' SED fits. 

Because our corrections are roughly the same, we will find the same result (SPIRE flux density vs. density), and so we do not take this error in the SFR into account. If we had picked the wrong SED, then all of our data points would move up or down together, leading to the same conclusion in whether or not the MS is dependent of environment. 

For the SFR-density in the COSMOS field we constructed the errors based on a tile-selected jackknife over the map, in combination with the error on the SFR of the stacked galaxies and the error of the mass function, see \cite{2013A&A...556A..55I}. In this case, we have to take the error in the template into account, because we want to compare with previous results for the SFR-density relation.  

We quantified this uncertainty by not only running the \texttt{SIMSTACK} code for the best SED template, but also for the second to fifth best templates. These different templates give different SFR estimates. We constructed a weighted mean of these SFRs by weighting each SFR by the reduced $\chi^2$ (on which we based our choice of best templates). 

By enforcing the reduced $\chi^2$ of this mean to be 1 we enlarge these errors. These enlarged errors based on our best templates give a better estimate of the uncertainty in the SFR-density by also including the bias from selecting a specific template for the SED.

For our environmentally dependent SFR we did not take this bias into account. In each of the chosen top five SEDs the same environmental trend can be seen as we observe for our best template; so by taking the bias in SFR into account for this plot, we will wash out any observed correlation. Therefore we can say that there is an extra uncertainty on the SFR estimates (as seen in Figure \ref{fig:density_2}), but our environmental results would already be seen in using the higher SPIRE flux densities fitted to the map, justifying the use of the smaller error bars in Figure \ref{fig:SFR}.

\label{lastpage}

\end{document}